# Ultrafast Optical Nonlinearity in Poly(methyl methacrylate)-TiO$_2$ Nanocomposites

H. I. Elim, and W. Ji[*]

Department of Physics, National University of Singapore, Singapore 117542

A.H. Yuwono, J.M. Xue, and J. Wang

Department of Materials Science, National University of Singapore, Singapore 117543

**Abstract**

*With 780-nm, 250-fs laser pulses, ultrafast optical nonlinearity has been observed in a series of thin films containing poly(methyl methacrylate) (PMMA)-TiO$_2$ nanocomposites, which are synthesized by a simple technique of in-situ sol-gel/polymerization. The best figures of merit are found in one of the films prepared with a 60% weight percentage of titanium isopropoxide. Transmission electron microscopy (TEM) shows the presence of the 5-nm-diameter particles in the film. The observed optical nonlinearity has a recovery time of ~1.5 ps. These findings suggest the strong potential of PMMA-TiO$_2$ nanocomposites for all-optical switching.*

---

[*] Electronic mail: phyjiwei@nus.edu.sg

Nano-scale composite materials containing titanium oxides are interesting because of their potential applications in optoelectronic devices.[1-2] A great deal of research effort has been focused on both synthesis of high-quality, transparent films consisting of polymer-$TiO_2$ hybrid nanocomposites, and their linear optical properties.[3-7] Recently, the nonlinear-optical properties of such materials have also received attention. The third-order optical nonlinearity ($\chi^{(3)} = 4.1$ x $10^{-11}$ esu) in $TiO_2$ nano-crystalline particles dispersed in $SiO_2$ was measured at 1.06 μm with nanosecond laser pulses.[8] Large, reverse-saturable type of nonlinear absorption was also observed in poly(styrene maleic anhydride)/$TiO_2$ nanocomposites with cw He-Ne laser beam.[9] Here we report our investigation into ultrafast optical nonlinearity in a series of thin films of poly(methyl methacrylate) (PMMA)-$TiO_2$ nanocomposites. Our results show that such nanocomposites possess very large and ultrafast optical nonlinearity, and have a great potential for optical switching and optical communications.

The thin films of PMMA- $TiO_2$ nanocomposites were prepared based on a modification of a reported method.[7] Monomers, methyl methacrylate (MMA) and 3-(trimethoxysilyl) propyl methacrylate (MSMA), and an initiator benzoyl peroxide (BPO) in tetrahydrofuran (THF) were added into reaction flask and polymerized at $60^{o}$C for 1 hour. The molar ratio of MSMA to MMA+MSMA was 0.25 and the amount of BPO that was added to this mixture was fixed at 3.75 mol%. On the other hand, a $TiO_2$-based solution was prepared using titanium isopropoxide (Ti-iP), de-ionized water, ethanol and hydrochloric acid, according to the method described elsewhere.[10] Ti-iP was first mixed with ethanol in a container and stirred for 30 minutes. A mixture of de-ionized water and HCl was poured under stirring into the transparent solution to promote hydrolysis. The Ti-iP concentration in the solution was fixed at 0.4 M with an understoechiometric ratio of water to Ti-iP (rw) of 0.82. And the pH value of 1.3 was used to obtain a stable solution (i.e. with longest gelation time). Furthermore, this homogeneous mixture was added dropwise over 30 minutes into the polymerized monomers with rigorous stirring to avoid local inhomogenities. The reaction was allowed to proceed at $60^oC$ for another 2 hours. Finally, five films according to the Ti-iP weight percentage in the preparation (i.e. 20, 40, 60, 80 and 100%, labeled as T20, T40, T60, T80 and T100, respectively,) were prepared by the following procedure. Each solution was spin-coated on quartz substrates at 3000 rpm for 20 seconds. Prior to the spin coating, the substrates were carefully cleaned, first in diluted $HNO_3$ solution in an ultrasound bath. After a thorough rinsing in running water, the ultrasound bath treatment was repeated with distilled water, acetone and ethanol. The substrates were then dried and stored in the drying oven at $40^oC$. The coated films were then baked in two stages of curing temperatures to finish the polymerization, i.e. at $60^oC$ for 30 minutes and $150^oC$ for 3 hours. The thickness of the films was measured to be varying from 250 to 350 nm by a surface profiler (Alpha-Step 500).

All the films are transparent to the human eye and their linear transmittance is 80~90% at 780 nm. Fig. 1(a) displays their $(\alpha h\nu)^{1/2}$ spectra, where $\alpha$ is the linear absorption coefficient and $h\nu$ is the photon energy. The absorption onsets change un-monotonically as the titanium content increases. As the weight percentage of Ti-iP increases from 20% to 60%, there is a red shift in the absorption, consistent with published reports.[4,5,7,9] The maximum red shift is found in the T60 film, which also has an absorption tail spanning the entire visible spectrum. It is due to the formation of nanoparticles. In Fig.1(b), the high resolution transmission electron microscopy (TEM) clearly shows the presence of the nanoparticles and Fig. 1(c) shows their size distribution with an average diameter of 5 nm. These particles give rise to loss through Rayleigh scattering, which is commonly associated with guest-host structures with different refractive indices for guest and host materials.[2] In addition, weak confinement may also contribute to the absorption below the band-gap energy. The exciton Bohr radius, $a_B$, in bulk TiO$_2$ solid is 0.8 nm if the effective masses ($m_e$ =10 $m_0$ and $m_h$ = 0.8 $m_0$) are used.[11] In the nanoparticles, the exciton energy can be estimated by using the Brus formula:[12] $E_{exc} = E_g + R_y$ ($\pi^2 a_B^2/R^2$-3.6 $a_B/R$), where $R_y$ is the Rydberg energy unit, $E_g$ (=3.2eV) is the band gap energy of anatase TiO$_2$ solid,[11] and $R$ is the particle radius. Two terms in the brackets represent kinetic energy and Coulomb interaction. With $R$ = ~2.5 nm, we find there is a red shift for the exciton energy, located ~3.16 eV. Note that our TEM evidences confirm much less or no formation of nanoparticles in the T80 and T100 films, which explains a reverse (blue) shift in the absorption as the titanium contents increase further. For the T100 film, the measured energy gap of 3.2 eV implies that the titanium oxides should be TiO$_2$ in the anatase crystalline phase.[11]

The nonlinear-optical properties of the PMMA- TiO$_2$ films were measured by Z-scan technique with 250 fs laser pulses at 780 nm. Figure 2(a) shows an example of open-aperture and closed-aperture Z-scans of the T60 film performed in two different repetition rates of 0.4 kHz and 4.0 MHz. The results show no significant difference between these Z-scans, indicating that laser-induced thermal lensing effects are negligible. The intensity independence of the Z-scans in Fig. 2(b) shows pure third-order nonlinear processes for the observed nonlinearities. Therefore, the nonlinear absorption and refraction can be described by $\Delta\alpha = \beta I$, and $\Delta n = n_2 I$, where $\beta$ and $n_2$ are the nonlinear absorption coefficient and nonlinear refractive index, respectively, and $I$ is the light intensity. Both $\beta$ and $n_2$ values can be extracted from the best fitting between the Z-scan theory and the data.[13]

Figure 2(c) displays both $\beta$ and $n_2$ values of the five films plotted as a function of the weight percentage of Ti-iP. These nonlinear parameters increase with increasing the titanium content in the T20, 40 and 60 films. This behavior is expected as the TEM studies show an increasing concentration of the nanoparticles. The T60 film possesses the highest nonlinear absorption, with $\beta = 1.4 \times 10^3$ cm/GW, or Im($\chi^{(3)}$) = 0.89 x $10^{-9}$ esu, which is about 100 times larger than that for a rutile crystal of $TiO_2$, measured at 532 nm.[14] We attribute this enhancement to the resonance with the exciton transition at ~3.16 eV since the two-photon energy of the laser pulses is 3.18 eV. Unfortunately, theoretical values of the two-photon absorption for $TiO_2$ nanoparticles are unavailable from literature. However, the two-photon absorption cross sections of ~ $10^{-46}$ $cm^3$ s have been calculated for two-photon-allowed transitions in CdSe nanocrystals in 2.9-nm diameter.[15] For comparison, our data show that the two-photon absorption cross sections are in the range from $10^{-46}$ to $10^{-45}$ $cm^3$ s for the PMMA-$TiO_2$ nanocomposites.

Similarly, the highest $n_2$ (= 2.5 x $10^{-2}$ $cm^2$/GW, or Re($\chi^{(3)}$) = 1.7 x $10^{-9}$ esu) is also found in the T60 film, about two-orders of the magnitude higher than that measured by Zhou *et al* on $TiO_2$/$SiO_2$ nanocomposites at 1.06 μm.[8] Such a large difference is anticipated due to the following reasons: (1) the dispersion of the nonlinear refraction, in particular, our measured value is enhanced by the two-photon resonance; (2) different size and volume fraction, in particular, Zhou's result was obtained from $TiO_2$/$SiO_2$ nanocomposites of larger sizes (20-50 nm), in which $\chi^{(3)}$ should be expected to approach to the bulk value; and (3) different host materials. The positive sign of the measured $n_2$ is in agreement with theories for a photon energy that is nearly half of the transition energy.[16]

The optical nonlinearities in the T100 film are too small to be detected accurately. Because this film is a layer of anatase $TiO_2$, as indicated by both TEM and absorption spectrum studies, we employ the two-band theory[16] to estimate $\beta$ and $n_2$. The estimated values are at least two orders of the magnitude smaller than those found in the T60 film, which are very close to, or below the sensitivity of our detection system. To evaluate the material requirements for all-optical switching devices, we calculate the following figures of merit (FOM) for each thin film,

$W = n_2 I / \alpha_0 \lambda$, and $T = \beta \lambda / n_2$.[17] For $\lambda = 780 \ nm$ and $I = 5.9 \ GW/cm^2$, the best figures of merit have been obtained from the T60 film, in which $W = 1.2$ and $T = 4.3$, close to the target values of $W > 1$ and $T < 1$.

To assess the response time of the observed nonlinearity in the $TiO_2$-PMMA thin films, we performed a degenerate femtosecond time-resolved pump-probe technique. Figure 3 displays transient nonlinear absorption signals of the films. Except for the T100 film, the temporal profile of each signal consists of two components, shown by the fits using two exponentially decay terms. The first is an instantaneous component, which is determined by the laser pulse duration (250 fs). Another one is a slowly decaying component with a characteristic time about 1.5 ps, with the accurate times depending on the weight percentage of Ti-iP. This slow decay is due to the relaxation of excitons excited by the two-photon absorption. The largest nonlinear absorption signal, $\left| \Delta T / T \right|$, found in the T60 film is consistent with its highest nonlinear absorption measured previously using the Z-scan method. The picosecond recovery time at room temperature suggests strongly potential applications of this material in optical switching devices.

In conclusion, large optical nonlinearities in the PMMA-$TiO_2$ nanocomposites have been observed using 780-nm, 250-fs laser pulses, and the nonlinear response time has been found to be about 1.5 ps. The best figures of merit have been found in the film using 60% of Ti-iP in preparation. The large optical nonlinearities are believed to be caused by two-photon-resonant exciton in the nano-sized particles containing $TiO_2$. Finally it should be pointed out that these films of the nanocomposites are fabricated with a simple sol-gel/polymerization method.

# References

[1] G.H. Du, Q. Chen, R.C. Che, Z.Y. Yuan and L.M. Peng, *Appl. Phys. Lett.* **79**, 3702 (2001).

[2] N. Suzuki, Y. Tomita and T. Kojima, *Appl. Phys. Lett.* **81**, 4121 (2002).

[3] B. Wang, G.L. Wilkes, J.C. Hedrick, S.C. Liptak and J.E. McGrath, *Macromolecules* **24**, 3449 (1991).

[4] J. Zhang, S. Luo and L. Gui, *J. Mater. Sci.* **32**, 1469 (1997).

[5] W.C. Chen, S.J. Lee, L.H. Lee and J.L. Lin, *J. Mater. Chem.* **9**, 2999 (1999).

[6] J. Zhang, B. Wang, X. Ju, T. Liu and T. Hu, *Polymer* **42**, 3697 (2001).

[7] L. Lee and W. C. Chen, *Chem. Mater.* **13**, 1137 (2001).

[8] Q.F. Zhou, Q.Q. Zhang, J.X. Zhang, L.Y. Zhang, and X. Yao, *Mater. Lett.* **31**, 41 (1997).

[9] S.X. Wang, L.D. Zhang, H. Su, Z.P. Zhang, G.H. Li, G.W. Meng, J. Zhang, Y.W. Wang, J. C. Fan and T. Gao, *Phys. Lett. A* **281**, 59 (2001).

[10] M. Burgos and M. Langlet, *J. of Sol-Gel. Sci. Tech.* **16**, 267 (1999).

[11] S. Monticone, R. Tufeu, A.V. Kanaev, E. Scolan and C. Sanchez, *Appl. Surf. Sci.* **162/163**, 565-567 (2000).

[12] L. E. Brus, *J. Chem. Phys.* **79**, 5566 (1983), or *J Chem. Phys.* **80**, 4403 (1984), or *J. Chem. Phys.* **90**, 2555 (1986).

[13] M. Sheik-Bahae, A.A. Said, T. Wei, D.J. Hagan, and E.W. Van Stryland, *IEEE J. Quantum Electron* **26**, 760 (1990).

[14] Y. Watanabe, M. Ohnishi and T. Tsuchiya, *Appl. Phys. Lett.* **66**, 3421 (1995).

[15] M.E. Schmidt, S.A. Blanton, M.A. Hines and P. Guyot-Sionnest, *Phys. Rev. B* **53**, 12629 (1996).

[16]For example, M. Sheik-Bahae, D.J. Hagan and E.W. Van Stryland, *Phys. Rev. Lett.* **65**, 96 (1990).

[17]A. Miller, K. R. Welford and B. Baino, *Nonlinear Optical Materials for Applications in Information Technology*, edited by Kluwer Academic Publishers, Netherlands, p.301 (1995).

**Figure captions:**

FIG. 1 (a) Absorption-photon-energy spectra for five PMMA- $TiO_2$ films prepared according to weight percentages of Ti-iP, i.e. 20, 40, 60, 80 and 100%, and labeled as T20, T60, T60, T80, T100, respectively. (b) High-resolution TEM of the T60 film. (c) Size distribution of the nanoparticles in the T60 film. The solid line is a lognormal fit to the size distribution

FIG. 2. (a) Open- and closed-aperture Z-scans performed with 780-nm, 250-fs laser pulses at two different repetition rates of 0.4 kHz (open squares and triangles) and 4 MHz (open circles and diamonds), respectively, on the T60 film. The input irradiance used is $I$ = 5 $GW/cm^2$ with the beam waist, $\omega_0$ = 13 μm. (b) Irradiance independence of the nonlinear absorption coefficient (β) and nonlinear refractive index ($n_2$) for the T60 film. (c) The β and $n_2$ values plotted as a function of the weight percentage of Ti-iP in PMMA.

FIG. 3. Transient nonlinear transmission responses of the PMMA-$TiO_2$ films performed at pump beam intensity, $I$ = 2.2 $GW/cm^2$. The solid lines are the best fits based on two exponentially decay terms.

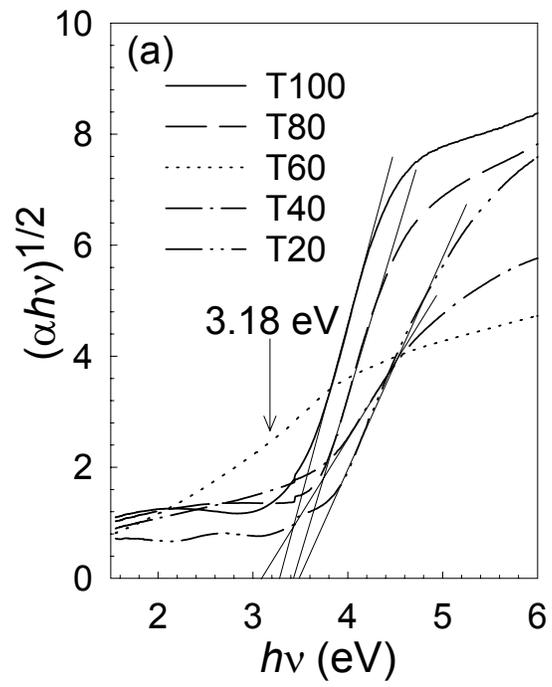

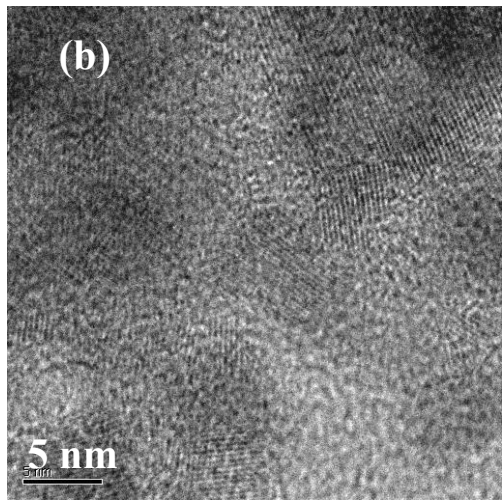

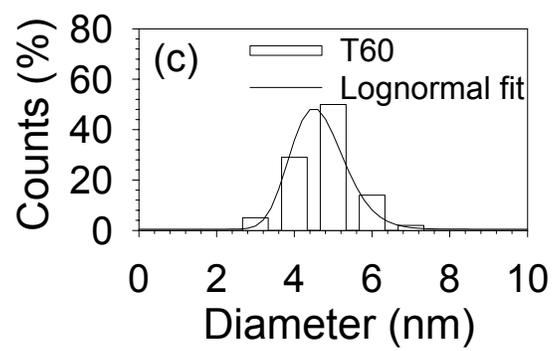

FIG. 1.

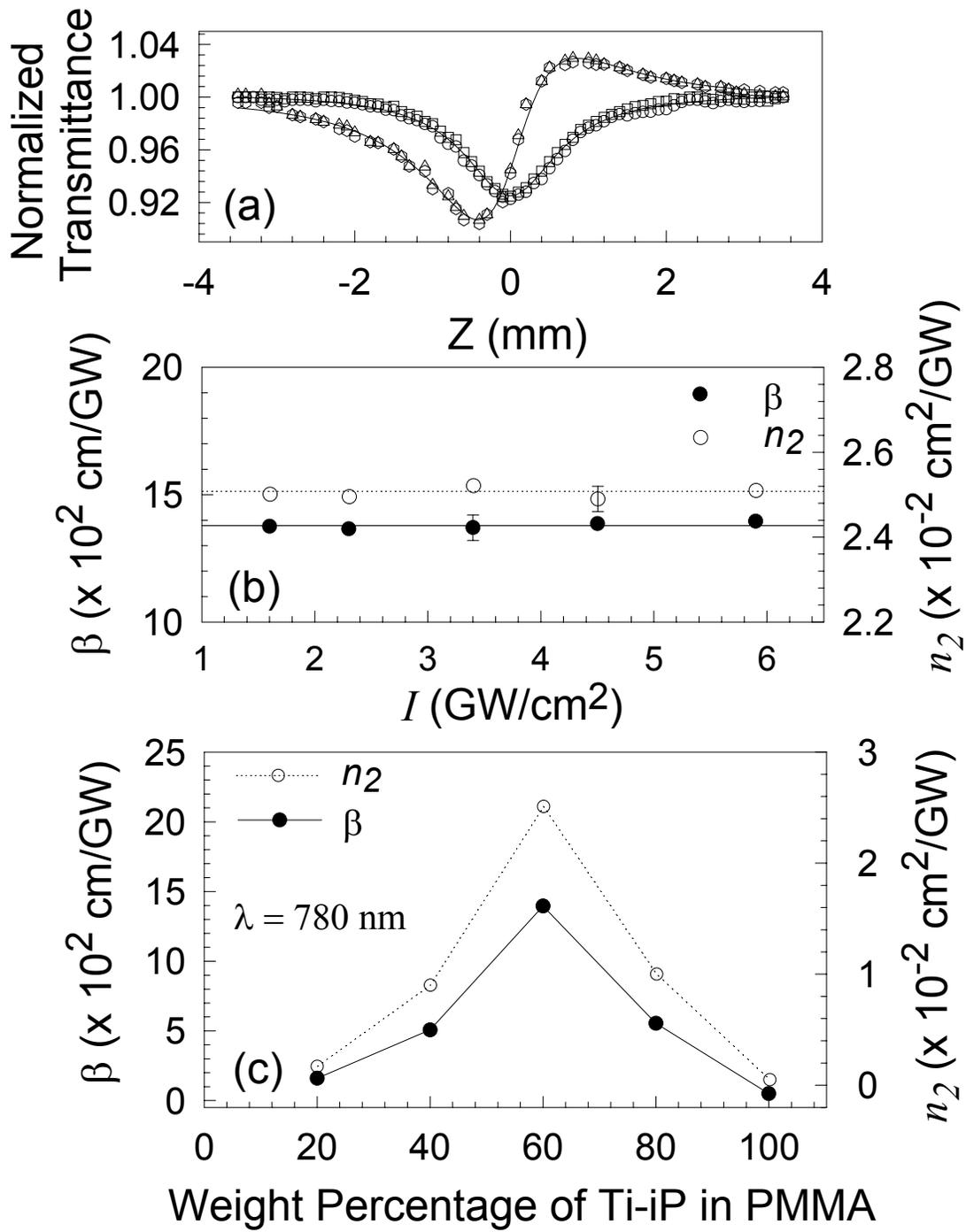

FIG. 2.

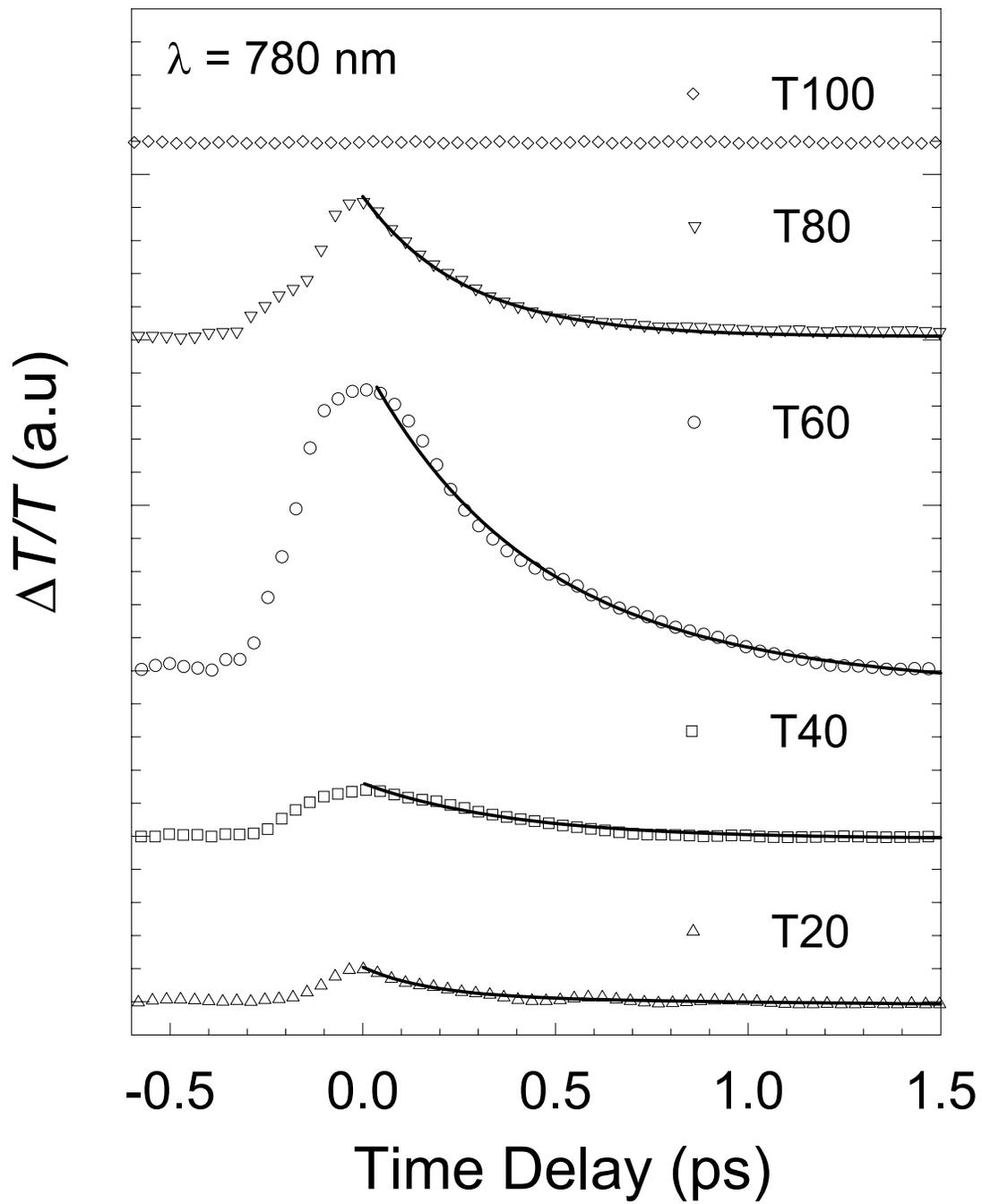

λ = 780 nm

T100
T80
T60
T40
T20

ΔT/T (a.u)

Time Delay (ps)

FIG. 3.